\documentclass{article}
\usepackage{spconf,amsmath,graphicx}
\usepackage{amsfonts}
\usepackage{cite}
\usepackage{multirow}
\usepackage{array}
\usepackage{ifpdf}
\usepackage{balance}
\title{Neural MOS Prediction for Synthesized Speech Using Multi-Task Learning With Spoofing Detection and Spoofing Type Classification}
\name{Yeunju Choi, Youngmoon Jung, Hoirin Kim\thanks{This material is based upon work supported by the Ministry of Trade, Industry \& Energy (MOTIE, Korea) under Industrial Technology Innovation Program (No. 10080667, Development of conversational speech synthesis technology to express emotion and personality of robots through sound source diversification).}}
\address{School of Electrical Engineering, KAIST, Daejeon, Republic of Korea}
\begin{document}
%\ninept
%
\maketitle
\begin{abstract}
Several studies have proposed deep-learning-based models to predict the mean opinion score (MOS) of synthesized speech, showing the possibility of replacing human raters. 
%수정
However, inter- and intra-rater variability in MOSs makes it hard to ensure the high performance of the models. 
%수정
In this paper, we propose a multi-task learning (MTL) method to improve the performance of a MOS prediction model using the following two auxiliary tasks: spoofing detection (SD) and spoofing type classification (STC).
Besides, we use the focal loss to maximize the synergy between SD and STC for MOS prediction. Experiments using the MOS evaluation results of the Voice Conversion Challenge 2018 show that proposed MTL with two auxiliary tasks improves MOS prediction.
Our proposed model achieves up to 11.6\% relative improvement in performance over the baseline model.
\end{abstract}
\section{Introduction}
\label{sec:intro}

Speech generation tasks such as text-to-speech and voice conversion have achieved great success in recent years with advances in deep learning \cite{WaveNet, Tacotron2, DeepVoice3, TransformerTTS, CycleGAN, NeuralTTS_VC}.
In terms of the quality (i.e., naturalness) of synthesized speech, state-of-the-art systems have reached human-level performance.
However, since there is no ``explicit answer'' for the speech generation tasks, researchers still rely on a subjective mean opinion score (MOS) test to evaluate the quality of the synthesized speech, which is expensive and time-consuming \cite{LimitOfMOS, Eval}. 
Although there are many objective measures of speech quality \cite{MCD, PESQ, ANIQUE, P563}, they have several limitations as discussed in our previous work \cite{Choi}.
For the MOS test, researchers need to perform the following steps: 1) employing a large enough number of human raters, 2) giving them a guideline and time to evaluate synthesized utterances, 3) compiling statistics of their answers (i.e., obtaining utterance-level MOSs by averaging the answers from different raters for each utterance and calculating a system-level MOS by averaging all the corresponding utterance-level MOSs).
Additionally, different studies may produce inconsistent results when evaluating the same speech generation system, due to the difference of test designs and human raters.

To address such problems of the subjective MOS test, researchers have recently proposed deep-learning-based models to predict the MOS of synthesized speech \cite{HierarchicalAssessment, AutoMOS, MOSNet, Choi}.
Yoshimura \textit{et al.} \cite{HierarchicalAssessment} proposed a hierarchical model that predicts the utterance-level and system-level MOS at once. 
Note that they statistically confirmed that the subjective MOSs of the dataset they used are inherently predictable.
Patton \textit{et al.} \cite{AutoMOS} proposed AutoMOS, based on long short-term memory (LSTM), to predict the MOS. 
Lo \textit{et al.} \cite{MOSNet} proposed MOSNet based on convolutional neural network-bidirectional LSTM (CNN-BLSTM), which produces an utterance-level MOS using frame-level scores.
Our previous work \cite{Choi} proposed two kinds of cluster-based modeling for learning the MOS evaluation criteria and utilizing the frame-level scores more precisely, respectively.

Even though these works have been successfully carried out to predict the MOS, inter- and intra-rater variability in MOSs remains a fundamental problem that limits the performance of MOS prediction models.
%수정
%To compensate for the unstable MOSs, we propose a multi-task learning (MTL) \cite{MTL_1993, MTL_overview} method for MOS prediction. MTL is a famous regularization technique that uses the information from related tasks, which are used as auxiliary tasks for the main task. 
%We expect that multi-task learning (MTL) \cite{MTL_1993, MTL_overview}, a famous regularization technique, can alleviate the problem. 
We expect that multi-task learning (MTL) \cite{MTL_1993, MTL_overview} can alleviate the problem by using the information from related tasks, which are used as auxiliary tasks for the main task. 
MTL is also known to improve the generalization ability of neural networks.
%by using the information from related tasks, which are used as auxiliary tasks for the main task. 
%MTL helps a model to generalize better by using the information from related tasks, which are used as auxiliary tasks for the main task. 
It has been successfully applied to various research areas \cite{MTL_NLP, MTL_SR, MTL_KD, MTL_SV_high_order, MTL_SV_noisy}. %MTL_CV, 

As suggested by \cite{BeneficialTask}, identifying beneficial auxiliary tasks is important in MTL. 
For example, for slot filling in natural language understanding, Louvan and Magnini \cite{NER_aux} proposed to use named entity recognition (NER) as an auxiliary task.
In this work, we propose an MTL method with spoofing detection (SD) and spoofing type classification (STC) to help MOS prediction.
To the best of our knowledge, this is the first work that uses MTL for MOS prediction. We also present a detailed analysis of the effects of our approach.
Besides, we use the focal loss \cite{FocalLoss} for SD to improve our MTL approach.
Experiments using the MOS evaluation results of the Voice Conversion Challenge (VCC) 2018 \cite{VCC2018} show that both auxiliary tasks help MOS prediction.
They also demonstrate that SD can create more synergy with STC by using the focal loss.

\begin{figure}[t]
    \centering
    \begin{small}
    \renewcommand{\arraystretch}{0.0}
    \includegraphics[trim=2.0cm 1.5cm 2.0cm 0.1cm, clip=true, width=8.6cm]{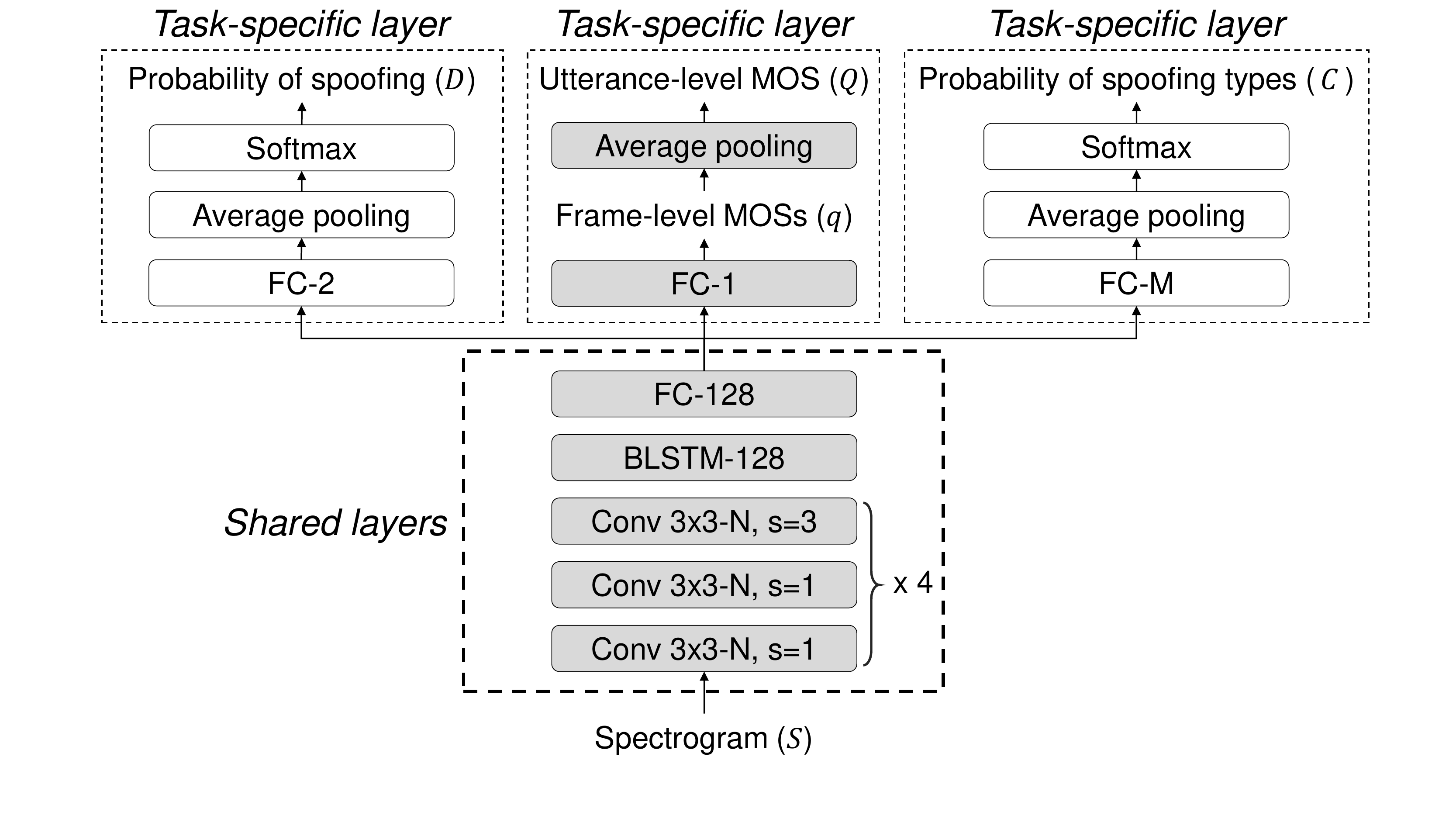}
    \end{small}
\vspace{-0.4cm}
\caption{
Overview of the proposed model.
The MOS prediction model is shown in gray. 
N is the number of channels for three convolutional layers, corresponding to 16, 16, 32, and 32 for four stacks.
M is the number of spoofing types and s is the stride for the convolutional layer.}
\label{fig:architecture}
\end{figure}

\section{Proposed methodology}
\label{sec:proposed_methodology}

\subsection{MOS prediction model}
\label{sec:mos_prediction_model}
%For the MOS prediction model, we adopt the recently proposed MOSNet \cite{MOSNet} that predicts the MOS of synthesized speech with a deep neural network (DNN) using a large open dataset, which consists of the MOS evaluation results of the VCC 2018 (VCC'18).
For the MOS prediction model, we adopt the recently proposed MOSNet \cite{MOSNet}. MOSNet predicts the MOS of synthesized speech with a deep neural network (DNN) using a large open dataset. Here, the dataset consists of the MOS evaluation results of the VCC 2018 (VCC'18).
In \cite{MOSNet}, Lo \textit{et al.} proposed different architectures for MOSNet: CNN, BLSTM, and CNN-BLSTM.
In this work, we use the CNN-BLSTM-based MOSNet as our baseline model since it achieved the best results among them.

The architecture of MOSNet is shown in gray in Fig. \ref{fig:architecture}.
The input and output of the model are a 257-dimensional magnitude spectrogram and a MOS, respectively, of an utterance.
From the input spectrogram, the CNN-BLSTM network extracts frame-level features.
The following two fully-connected (FC) layers predict frame-level MOSs from the frame-level features.
Here, each frame-level feature vector is used for its corresponding frame-level MOS value.
Then the frame-level scores are aggregated into an utterance-level MOS by average pooling.
Note that both frame- and utterance-level MOSs are scalar values.

The overall loss function is as follows:
\begin{equation}
L = \frac{1}{U}\sum_{u=1}^U[(\hat{Q}_u-Q_u)^2+\frac{\alpha_f}{T_u}\sum_{t=1}^{T_u}(\hat{Q}_u-q_{u, t})^2],
\end{equation}
where the first and second terms are mean squared errors (MSEs) for the utterance- and frame-level MOSs, respectively. $u$ is an utterance index, and $U$ is the number of utterances. $\hat{Q}_u$ and $Q_u$ are the ground-truth MOS and predicted MOS for the $u$-th utterance, respectively. $t$ is a frame index, and $T_u$ is the length of the $u$-th utterance. $q_{u, t}$ is the predicted frame-level MOS at the $t$-th frame of the $u$-th utterance. $\alpha_f$ is a loss weight for frame-level MOS prediction.

\subsection{Multi-task learning setup}
\label{sec:mtl_setup}
To improve MOS prediction, we propose MTL with two auxiliary tasks: spoofing detection (SD) and spoofing type classification (STC). 
In this work, SD indicates a binary classification to classify human speech as ``human" and synthesized speech as ``spoofing" \cite{SSD}.
STC is a multi-class classification to identify the source of input speech, which can be a speech generation system or a human speaker.
As the synthesized speech can be used for spoofing, we call the speech generation system a ``spoofing system," and all the spoofing systems and human speakers are collectively called ``spoofing types."

Fig. \ref{fig:architecture} shows the architecture of our MTL model. The CNN-BLSTM network and the FC layer with 128 nodes (FC-128) are shared by all the tasks and represented as ``shared layers."
For MOS prediction, the FC-1 layer and the following average pooling layer are used to predict frame-level MOSs and an utterance-level MOS, respectively. 
For each auxiliary task, we assign an additional task-specific layer, 
which consists of FC, average pooling, and softmax layers. 
The number of nodes of the FC layer is the same as the number of classes for each task, i.e., 2 for SD and M for STC.
Here, M is the number of spoofing types, which is equal to the number of spoofing systems plus the number of human speakers.
%Here, M is the number of spoofing types that is equal to the number of spoofing systems plus 2 (corresponding to the source speaker and target speaker).

We use the cross-entropy loss for both auxiliary tasks.
Then we define the final loss function as follows:
\begin{multline}
L = \frac{1}{U}\sum_{u=1}^U[\alpha_m(\hat{Q}_u-Q_u)^2+\frac{\alpha_f}{T_u}\sum_{t=1}^{T_u}(\hat{Q}_u-q_{u, t})^2 \\ - \alpha_d \sum_{i=1}^2\hat{D}_{u, i}\log D_{u, i} - \alpha_c \sum_{j=1}^M\hat{C}_{u, j}\log C_{u, j}],
\end{multline}
where $\hat{D}_{u,i}$ and $D_{u,i}$ are the $i$-th dimension of the ground-truth and predicted probability of spoofing for the $u$-th utterance, respectively. $\hat{C}_{u,j}$ and $C_{u,j}$ are the ground-truth and predicted probability of $j$-th spoofing type for the $u$-th utterance, respectively. %M is the number of spoofing types.
$\alpha_m$, $\alpha_d$, and $\alpha_c$ are the loss weights for utterance-level MOS prediction, SD, and STC, respectively.
In this setup, the number of parameters only increases by 1.44\% (from 359k to 364k).

\begin{figure}[t]
    \centering
    \begin{small}
    \renewcommand{\tabcolsep}{0.5mm}
    \renewcommand{\arraystretch}{0.9}
        \begin{tabular}{cc}
            \includegraphics[trim=0.4cm 0.1cm 0.1cm 0.4cm, clip=true, width=4.15cm]{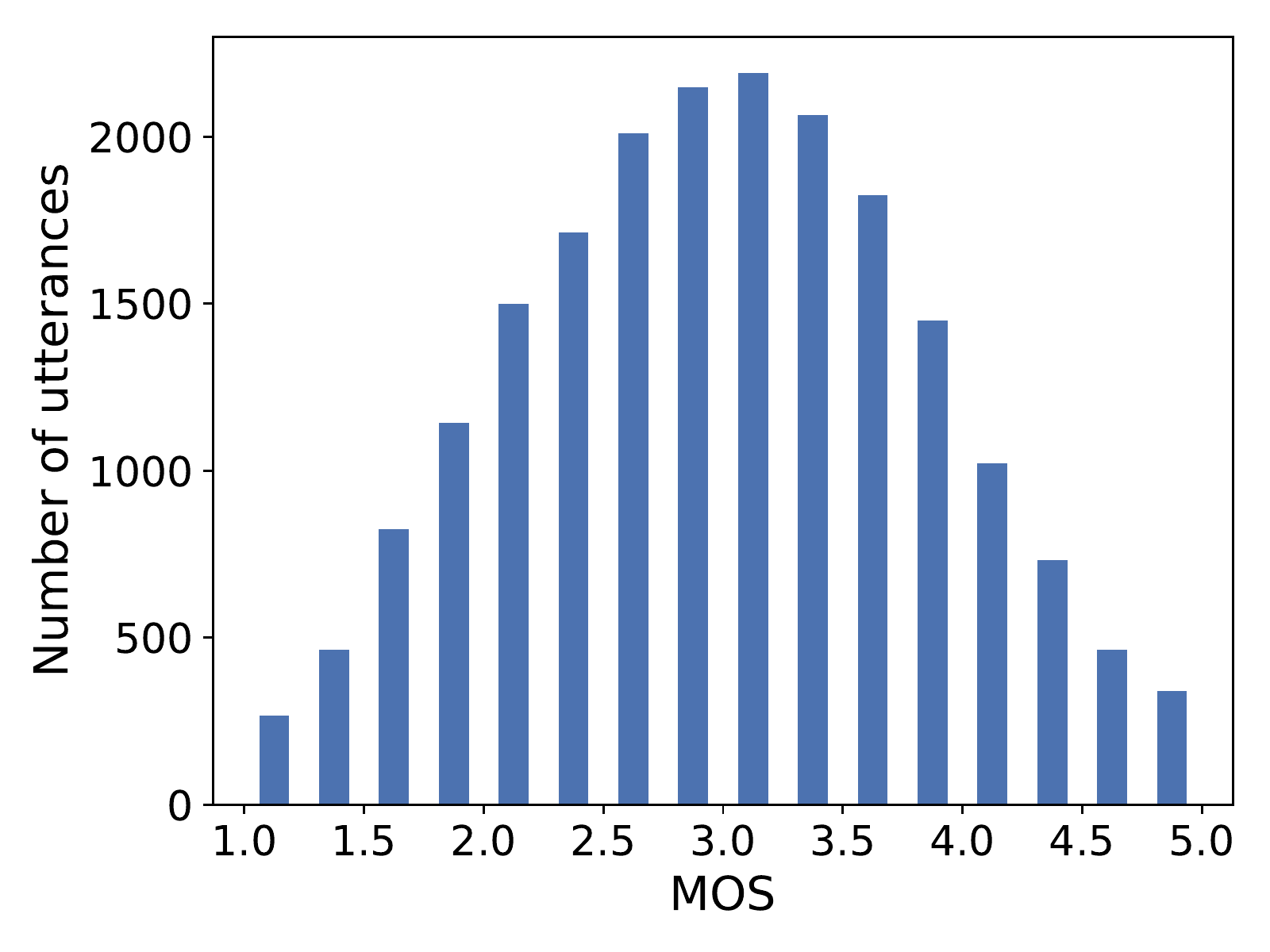} &
            \includegraphics[trim=0.05cm 0.1cm 0.45cm 0.4cm, clip=true, width=4.15cm]{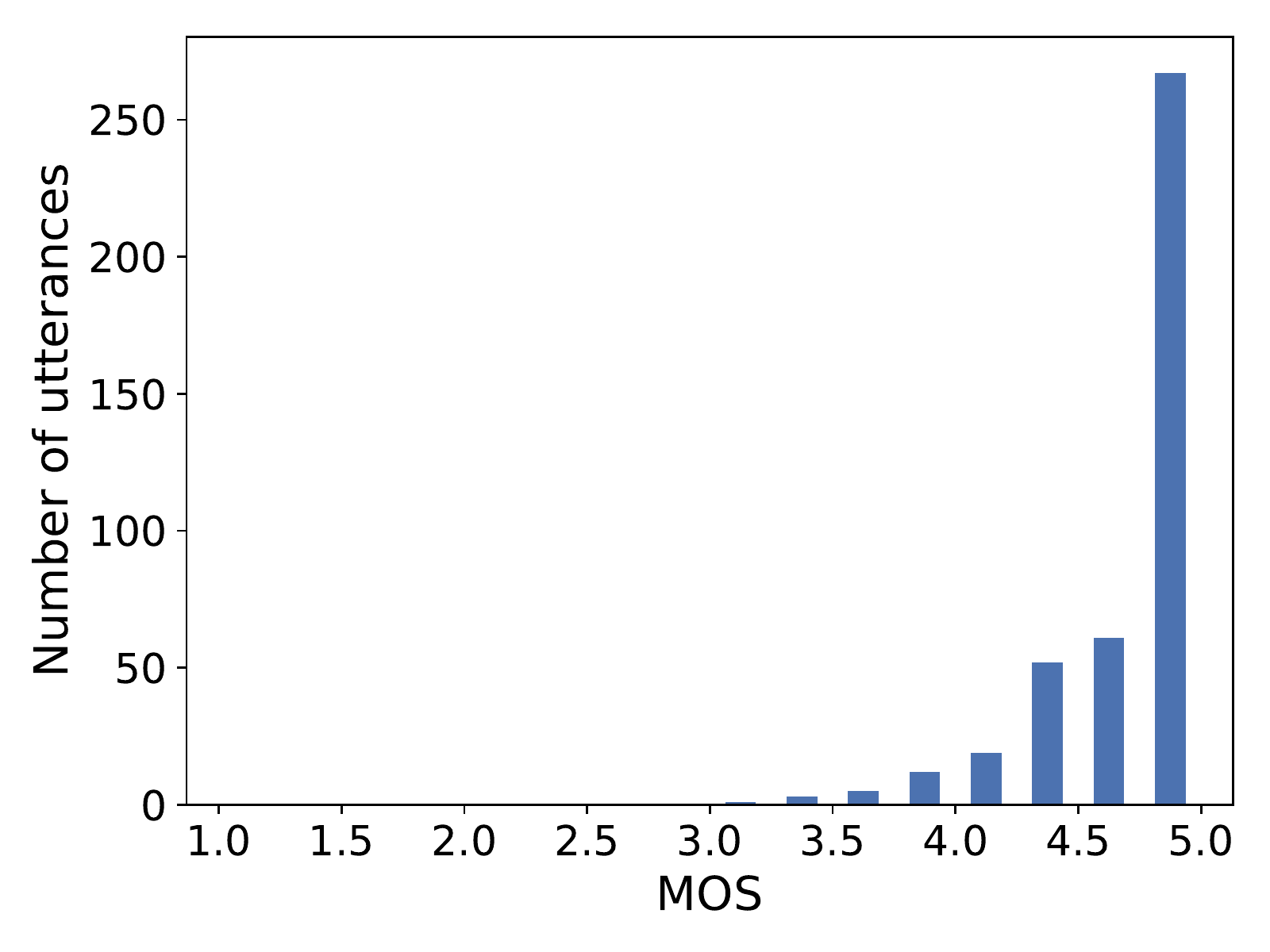} \\
            (a) Synthesized speech & (b) Human speech
        \end{tabular}
    \end{small}
\caption{Histograms of ground-truth utterance-level MOSs for (a) synthesized and (b) human speech from the VCC'18 data.}
\label{fig:VCC2018_dist}
\end{figure}

\subsection{Analysis of auxiliary tasks for MOS prediction}
\label{sec:auxiliary}

In this section, we explain how the two proposed auxiliary tasks can benefit MOS prediction. 
Before discussing the SD task, we first introduce the difficulty of predicting high MOSs.
Fig. \ref{fig:VCC2018_dist} shows the histograms of ground-truth utterance-level MOSs for the VCC'18 data.
We can see that there are much fewer high-quality (i.e., MOS $\ge$ 4.0) utterances than the others among the synthesized utterances.
%We can see that there are much fewer high-quality (i.e., having MOSs equal to or higher than 4.0) utterances than the others among the synthesized utterances.
%We can see that there are much fewer utterances with high MOSs (i.e., high-quality speech) than those with low MOSs (i.e., low-quality speech) among the synthesized utterances.
%Furthermore, the human speech, which usually has a high MOS, has a little portion in the total dataset since the purpose of the dataset is to evaluate speech generation systems.
Furthermore, human speech samples, which typically have high MOSs, constitute only a small fraction of the entire dataset. This is obvious since the purpose of the dataset is to evaluate synthesized speech rather than human speech.
Due to the lack of high-quality speech in the dataset, it is difficult for a MOS prediction model to predict high MOSs.
%In the next paragraph, we show that using the SD task can alleviate this problem. 
In the next paragraph, we explain how using the SD task can alleviate this problem. 

\begin{figure}[t]
    \centering
    \begin{small}
    \renewcommand{\arraystretch}{0.0}
    \includegraphics[trim=0cm 0.0cm 0.0cm 0.0cm, clip=true, width=8.6cm]{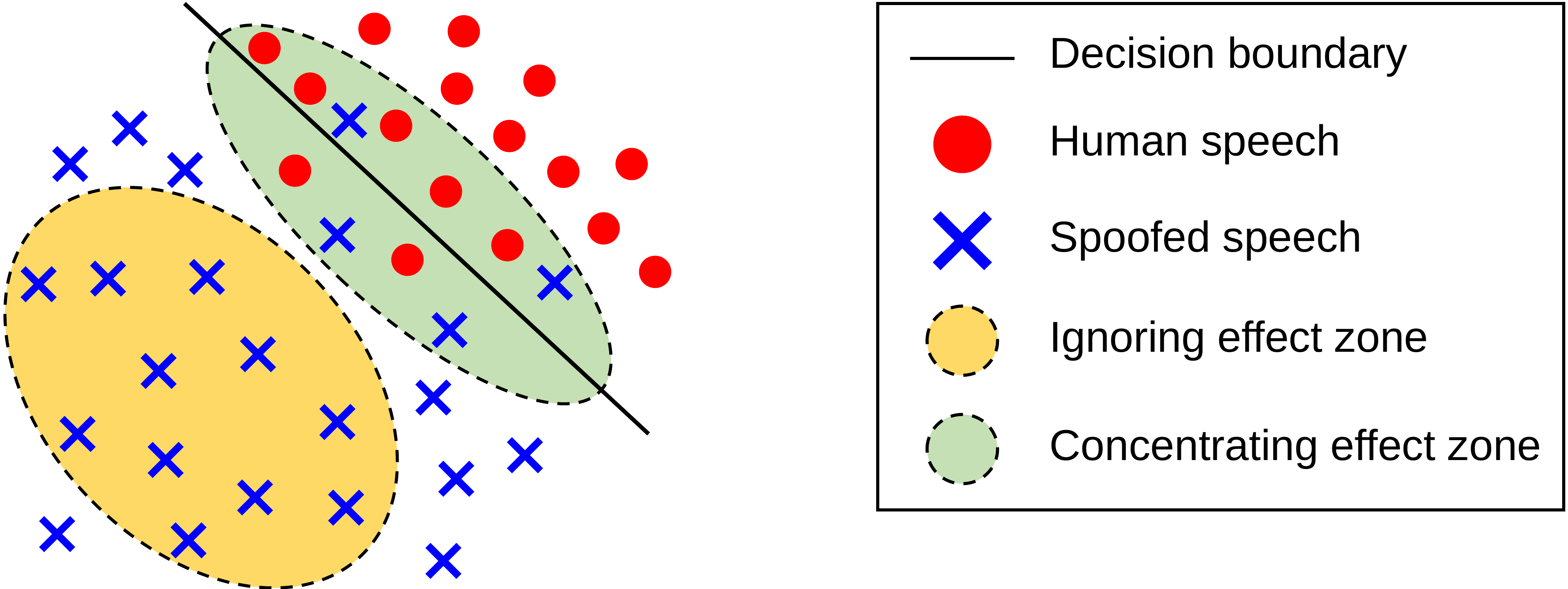}
    \end{small}
\caption{A conceptual illustration of a decision boundary for SD.}
\label{fig:decision_boundary}
\end{figure}

Fig. \ref{fig:decision_boundary} is a conceptual illustration of the decision boundary for SD that separates human speech and spoofed (or synthesized) speech. The data points correspond to the shared features extracted from the shared layers. 
From now on, we use this figure for a detailed analysis of SD.
For SD, the model is trained to discriminate between human speech and synthesized speech. 
Then it learns a decision boundary between the two classes, which is located in ``Concentrating effect zone" in Fig. \ref{fig:decision_boundary}.
Note that both human speech (marked as red circles) and high-quality synthesized speech (marked as blue `x' marks) exist in this zone.
Thus, if we train a model on SD, the model can learn to distinguish between the human speech and high-quality synthesized speech by giving attention to high-quality speech.
When trained on both MOS prediction and SD tasks, the model gives more attention to high-quality speech compared to when trained only on MOS prediction.
As a result, in MOS prediction, the model will predict high MOSs better by concentrating on the utterances with high MOSs. 
Accordingly, we call the effect above \emph{concentrating effect}.

As discussed in Section \ref{sec:mtl_setup}, for STC, the model is trained to identify the source of speech, called the spoofing type. 
Along with linguistic contents, a spoofing type determines low-level features of an utterance.
Therefore, the model learns the low-level features that are useful for distinguishing between various spoofing types.
% 고마와 스윗아가!! Note that low-level features indicate the extracted features that contain low-level information.
Here, the low-level feature can be any acoustic-prosodic feature that fundamentally affects the way of vocalization and intonation of spoofing types, e.g., fundamental frequency, intensity, duration, jitter, and shimmer. 
% pitch (F0-mean), intonation (F0-contour), intensity, phrasing (pattern of rate and pausing withing utterances), phoneme duration, rhythm, stress (emphasis on syllables and words), accentuation (amplitude envelope), noise-to-harmonics ratio
Considering that the combination of such low-level features determines the speech quality, we expect that the model can predict the MOS better by also learning to classify spoofing types.

Besides, the labels of both the SD and STC tasks are independent of the human raters. We believe such independence can further help to address the inter- and intra- rater variability problem in MOS prediction.
Experimental results in Section \ref{sec:results} support the effectiveness of using these auxiliary tasks.

\subsection{Focal loss for the spoofing detection task}
\label{sec:FL_SD}
In this section, we describe our motivation to use the focal loss for SD.
The SD model is trained to classify synthesized utterances into the same class, ``spoofing," even though the utterances are generated by various spoofing systems.
Then MTL with SD would prevent the model from distinguishing between spoofing systems.
We call this effect of SD \emph{ignoring effect} as it ignores the difference between spoofing systems. %unique 

However, when we train a MOS prediction model with STC, the model will be able to distinguish between the spoofing types in training data, which we call \emph{distinguishing effect}. Then, when testing the model using the same spoofing types in training data, using SD as another auxiliary task causes a conflict between the \emph{ignoring effect} and \emph{distinguishing effect}.
To maximize the synergy between SD and STC, we want to prevent SD from having the \emph{ignoring effect}. 
Note that there are much more synthesized utterances far from the decision boundary for SD, having low MOSs and SD losses, than those near the decision boundary, having high MOSs and SD losses (see Fig. \ref{fig:decision_boundary}). 
Those low-quality (i.e., MOS $<$ 3.0) synthesized utterances having low SD losses, located in ``Ignoring effect zone," contribute to only the \emph{ignoring effect} but not the \emph{concentrating effect}.
To reduce the contribution of the utterances with low SD losses, we adopt the focal loss for SD. % in Fig. \ref{fig:decision_boundary} \cite{FocalLoss} 

% Focal Loss formulation
The cross-entropy (CE) loss of the $u$-th utterance for the SD task is as follows:
\begin{equation}
CE(D_u, \hat{D}_u) = \left\{ 
\begin{array}{@{}ll@{}}
-\log D_{u,1}, & \text{if}\ \hat{D}_{u}=(1,0) \\ 
-\log D_{u,2}, & \text{if}\ \hat{D}_{u}=(0,1).
\end{array}\right.
\end{equation}
The first and second dimensions of $\hat{D}_u$ corresponds to the ``human" and ``spoofing" classes, respectively.
Please note that $D_{u,2} = 1 - D_{u,1}$ since we use the softmax layer.

Now we define $p_u \in [0, 1]$ as follows for convenience:
\begin{equation}
p_u = \left\{ 
\begin{array}{@{}ll@{}}
D_{u,1}, & \text{if the}\ u\text{-th utterance} \in \text{``human."} \\ 
1-D_{u,1}, & \text{otherwise.}
\end{array}\right.
\end{equation}
Then the CE loss can be rewritten as $CE(p_u) = -\log(p_u)$. To down-weight the contribution of easy samples (that have lower loss values) and focus on hard samples (that have higher loss values), the focal loss (FL) is defined as $FL(p_u) = -(1-p_u)^\gamma \log(p_u)$ with a parameter $\gamma \geq 0$ that adjusts the rate at which easy samples are down-weighted.

\section{Experiments}
\label{sec:experiments}

\subsection{Dataset}
\label{sec:dataset}
We use the MOS evaluation results of the VCC 2018 (VCC'18) \cite{VCC2018}, which is a large-scale, open, and intrinsically-predictable dataset.
%We denote this dataset as VCC'18 dataset.
A total of 38 systems participated in the VCC 2018 (i.e., M = 38), including two human speakers (source and target speakers).
The ground-truth MOS of a system was obtained by averaging the ground-truth MOSs of all the utterances from the system.
A total of 267 people rated a total of 20,580 utterances on a scale from 1 (completely unnatural) to 5 (completely natural).
An average of 4 people per utterance participated in the evaluation, leading to a total of 82,304 MOS evaluation results.
The ground-truth of the utterance-level MOS was obtained as the average of all the MOSs of the utterance.
As can be seen in Fig. \ref{fig:VCC2018_dist},
the MOS of synthesized speech almost follows a Gaussian distribution with the mean value near 3.0, while approximately two-thirds of human speech has a MOS value larger than 4.5. As mentioned in Section \ref{sec:auxiliary}, high-quality utterances are much fewer than the others in the total data.
From 20,580 $<$audio, ground-truth MOS$>$ pairs, we randomly select 15,580, 3,000, and 2,000 pairs for training, validation, and testing, respectively. For a more detailed explanation about the dataset, see \cite{Choi}.

In \cite{MOSNet}, Lo \textit{et al.} thoroughly discussed the suitability of the dataset for MOS prediction in terms of that the dataset is intrinsically predictable. 
According to their experiments using the bootstrap method \cite{bootstrap}, both the linear correlation coefficient (LCC) \cite{Pearson} and Spearman's rank correlation coefficient (SRCC) \cite{Spearman} of the dataset are larger than 0.80 and 0.97 at the utterance level and system level, respectively.
The experimental results suggest two conclusions: 1) the MOS is intrinsically predictable at the system level, and 2) the utterance-level MOS is predictable to some extent.
Moreover, MOSNet trained on the VCC'18 dataset showed the generalization ability to the MOS evaluation results from the VCC 2016 (VCC'16) \cite{VCC2016}. This proves the generalization ability of not only MOSNet but also the VCC'18 data.

We also use VCC'16 to evaluate the robustness of the models to unseen spoofing types and raters.
Including a target speaker, source speaker, and baseline system, a total of 20 systems exist in VCC'16.
For each system, 1,600 utterance-level MOS evaluation results exist without specification of the utterances or raters. 
Therefore, it is unable to use the utterance-level MOSs, and we only use the system-level MOSs with a total of 26,028 utterances.

\subsection{Implementation details}
\label{sec:imp_details}
All the models are implemented using PyTorch and trained on a single GTX 1080 Ti GPU. 
We use a batch size of 32 and the Adam optimizer with a learning rate of $10^{-4}$ for all the models.
We set the weights for utterance- and frame-level MOS prediction to 1 and 0.8, respectively.
For MTL, we set the weight of the loss for each auxiliary task to 1.
When we adopt the focal loss (FL) for the SD task, we set $\gamma$ to 0.8.

For testing, we use the model that has the lowest MSE on the VCC'18 validation set during 200 epochs of training.
Note that we train all the models using only the VCC'18 training set and test them on both the VCC'18 test set and VCC'16 data.
We conduct the experiments for each model with four different random seeds and report the average value of the four results as the performance of each model. 
The performance is evaluated in terms of the MSE, LCC, and SRCC.
We report both utterance- and system-level performances for the VCC'18 test set. For the VCC'16 data, we report only system-level performance because the utterance-level MOSs for VCC'16 are not available, as mentioned in Section \ref{sec:dataset}.

\begin{table*}[t]
\caption{Results of an ablation study and comparison with a state-of-the-art method. 
MOS and F-MOS are utterance- and frame-level MOSs, respectively. 
RI is the average relative improvement of nine metrics. 
MOSNet+EL is a state-of-the-art model, which is MOSNet using the Encoding Layer.
The best results are shown in bold.}
%\vspace{0.1cm}
\label{tab:mos_results}
\begin{small}
\begin{center}
\renewcommand{\tabcolsep}{1.1mm}
\renewcommand{\arraystretch}{1.05}
\begin{tabular}{l|c|cccc|cccccc|ccc|c}
\hline
\multicolumn{1}{c|}{\multirow{3}{*}{Model}} & {\multirow{3}{*}{FL}} & \multicolumn{4}{c|}{Loss weights}                   & \multicolumn{6}{c|}{VCC'18}                                            & \multicolumn{3}{c|}{VCC'16}   &  \multirow{3}{*}{RI (\%)} %\\
\tabularnewline\cline{3-15}
\multicolumn{1}{c|}{}          &             &  $\alpha_m$  &   $\alpha_f$  & $\alpha_d$  &     $\alpha_c$    & \multicolumn{3}{c}{\textit{utterance-level}} & \multicolumn{3}{c|}{\textit{system-level}} & \multicolumn{3}{c|}{\textit{system-level}} & \\
\multicolumn{1}{c|}{}                       &  & (MOS) & (F-MOS)      & (SD)      & (STC) &    MSE        & LCC       & SRCC       & MSE       & LCC      & SRCC      & MSE       & LCC       & SRCC     &  \\\hline\hline
Baseline                                   & -                            & 1             & 0.8            & 0       & 0       &   0.448   &  0.651   &  0.619    &  0.039   &  0.966  &  0.924   &  0.316   &  0.896   &  0.858  &  - \\
+SD                                   & -                            & 1             & 0.8            & 1       & 0       &    0.439  &  0.661   &  0.623    &  0.029   &  0.972  &   0.925  &  0.333   &  0.886   &  0.829  & 2.34\\
+SD                                   & \checkmark                            & 1             & 0.8              & 1       & 0       &    0.445  &  0.655   &  0.621    &  0.029   &  0.971  &  0.934   &  0.283   &  0.907   &  0.847  & 4.31\\
+STC                                  & -                            & 1             & 0.8            & 0       & 1       &   0.434   &  0.666   &  \textbf{0.625}    &  0.020   &  0.982  &  \textbf{0.955}   &  0.226   &  0.915   &  0.859  & 10.2\\
+STC +SD                               & -                            & 1             & 0.8            & 1       & 1       &   0.435   & 0.664    &   0.618   &  0.019   &  0.983  &  0.944   &  0.227   &  \textbf{0.925}   &  \textbf{0.883}  & 10.5 \\
+STC +SD                               & \checkmark                            & 1             & 0.8              & 1       & 1       &   \textbf{0.431}   &  \textbf{0.668}   &  0.622    & \textbf{0.016}    & \textbf{0.985}   &  0.944   &  \textbf{0.208}   & 0.904    &  0.864  & \textbf{11.6}
\\\hline
MOSNet+EL \cite{Choi}             & -           & 1             & 0.8            & 0       & 0       &    0.444  &  0.656   &  0.617    &  0.031   &  0.974  &   0.938  &  0.242   &  0.908   &  0.855  & 5.40\\\hline
\end{tabular}
%\vspace{-0.2cm}
\end{center}
\end{small}
\end{table*}

\section{Results and Discussion}
\label{sec:results}
\subsection{Effectiveness of auxiliary tasks}
\label{sec:auxiliary_tasks}

The results of an ablation study, presented in Table \ref{tab:mos_results}, show the effectiveness of the proposed MTL approach.
%The results of an ablation study to show the effectiveness of the proposed MTL approach are presented in Table \ref{tab:mos_results}.
%We discuss the effectiveness of the proposed MTL approach through an ablation study in Table \ref{tab:mos_results}.
%Table \ref{tab:mos_results} presents the results of the ablation study to show the effectiveness of the proposed MTL approach.
%We first discuss the effectiveness of two auxiliary tasks based on the results of MTL with SD (+SD) and MTL with STC (+STC) on VCC'18, which are in the second and fourth rows, respectively. 
We first discuss the effectiveness of two auxiliary tasks based on the results of MTL with SD (+SD), MTL with STC (+STC), and MTL with both of them (+STC +SD) on VCC'18, which are in the second, fourth, and fifth rows, respectively. 
+SD improves the baseline in terms of all the metrics due to \emph{concentrating effect}. %, as expected in Section \ref{sec:auxiliary}.
Here, we need to point out the \emph{ignoring effect}; +SD ignores the difference between spoofing systems to classify all the synthesized utterances into the same class, ``spoofing," and thus loses low-level information that is useful for spoofing type classification. 
%The low-level information is also important for MOS prediction, and thus the \emph{ignoring effect} can be a disadvantage for MOS prediction.
%The low-level information is also important for MOS prediction, as mentioned in Section \ref{sec:auxiliary}. 
The low-level information is also useful for MOS prediction, as mentioned in Section \ref{sec:auxiliary}. 
From that point of view, the \emph{ignoring effect} can be a disadvantage for MOS prediction.
However, the task-specific layer for MOS prediction learns to handle the deficiency of low-level information caused by the \emph{ignoring effect}, based on the VCC'18 training set. 
Therefore, for the VCC'18 test set that has the same spoofing types and overlapping human raters with the training data, the \emph{ignoring effect} does not harm the performance of +SD.

+STC also achieves better performance compared to the baseline.
+STC yields better results than +SD, which corresponds to the intuition that the model learns more information with a multi-class classification task (i.e., STC) than with a simple binary classification task (i.e., SD).
% added
More specifically, we can infer that learning low-level information is more important than concentrating on high-quality speech for MOS prediction.
The MTL model with STC and SD (+STC +SD) improves the performance of the baseline in terms of all the metrics except the utterance-level SRCC, which is almost the same as that of the baseline.
However, the performance is not better than that of +STC due to the conflict between the \emph{ignoring effect} and \emph{distinguishing effect}. 
In Section \ref{sec:visulization}, we provide further analysis of these effects through visualization.

\begin{figure}[t]
    \centering
    \begin{small}
    \renewcommand{\tabcolsep}{1mm}
    \renewcommand{\arraystretch}{0.8}
        \begin{tabular}{cc}
            \includegraphics[trim=0.4cm 0.22cm 0.1cm 0.41cm, clip=true, width=4.1cm]{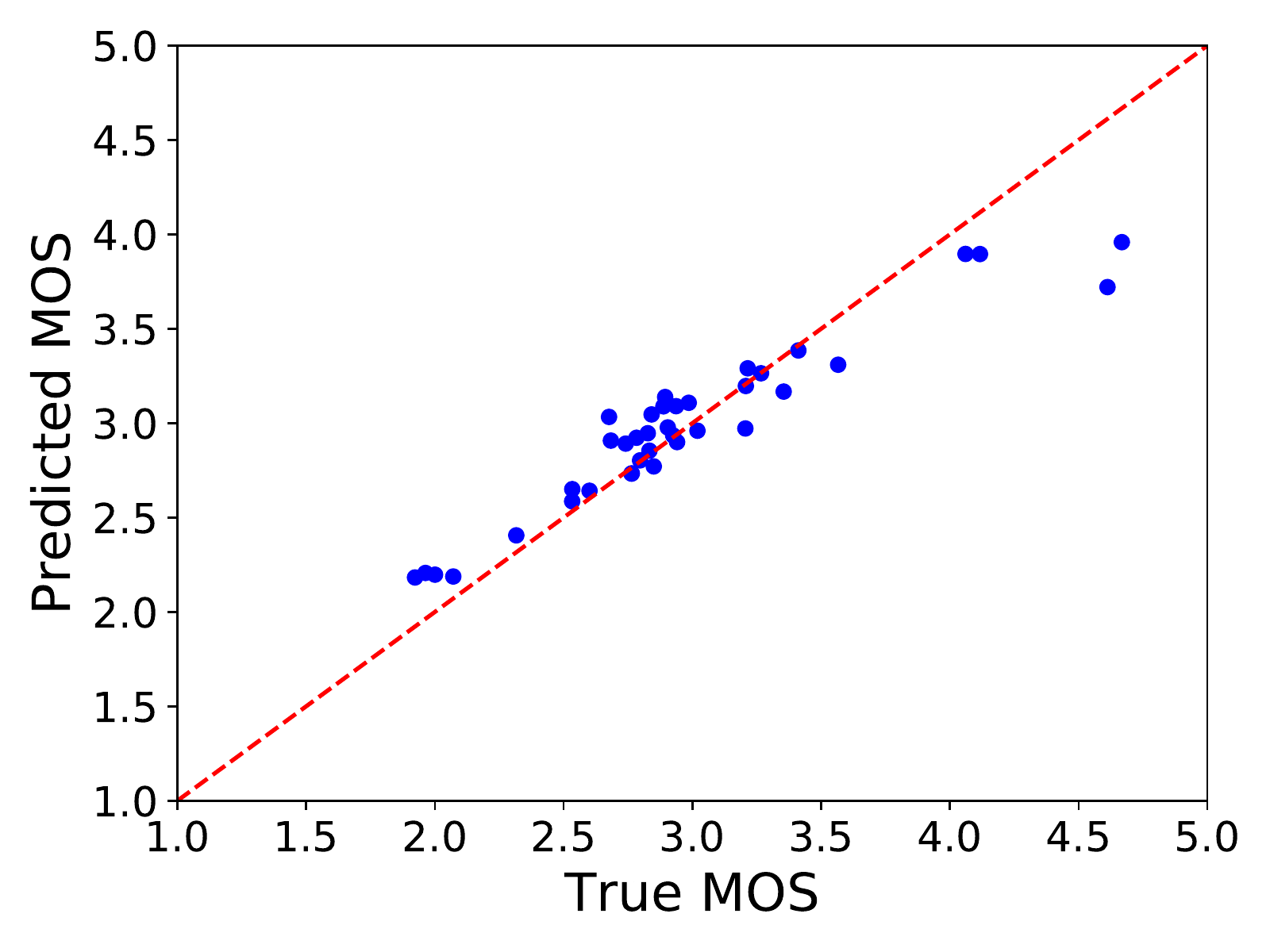} &
            \includegraphics[trim=0.2cm 0.22cm 0.3cm 0.41cm, clip=true, width=4.1cm]{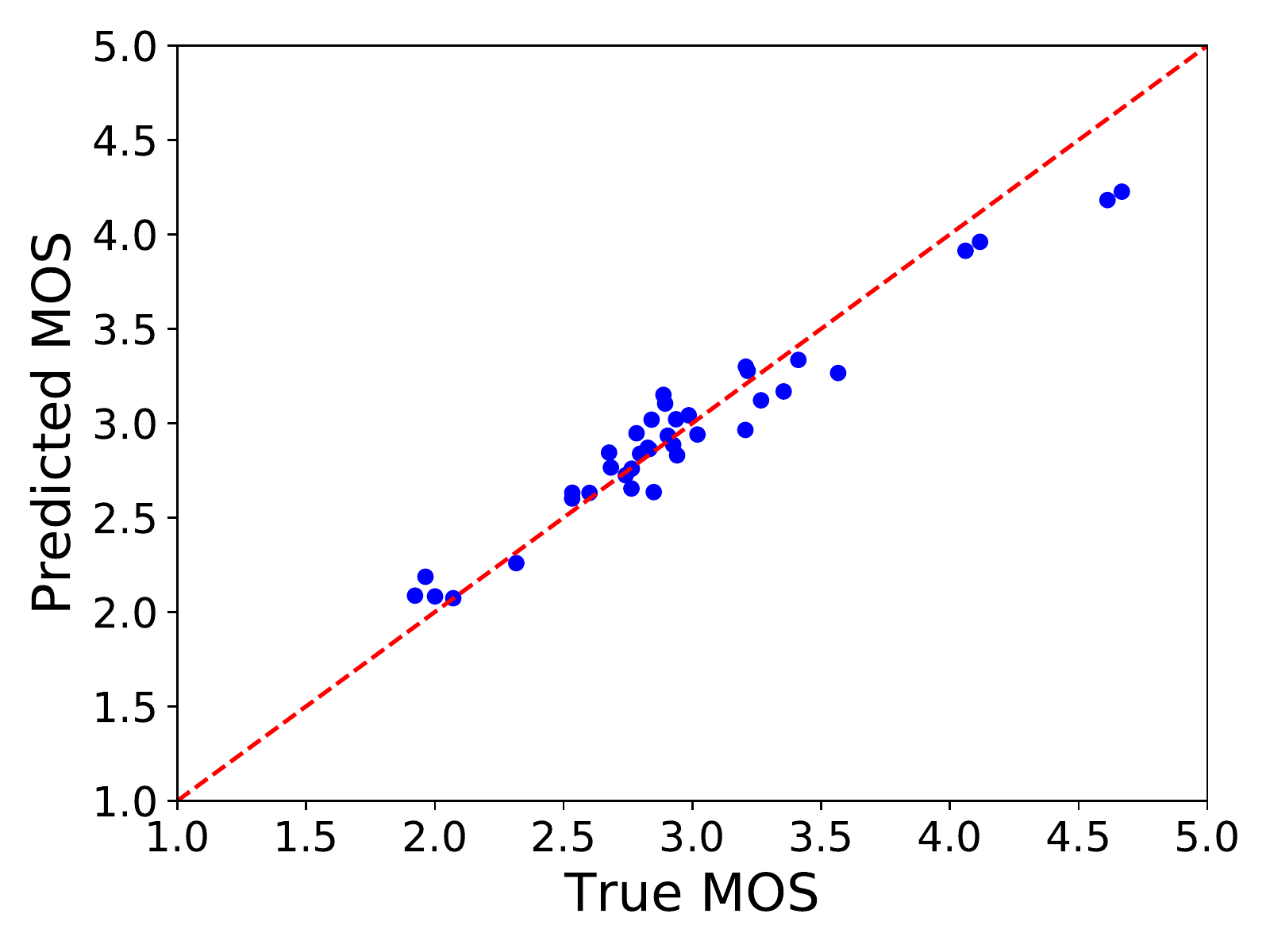} \\
            (a) Baseline & (b) +SD\\\\
            \includegraphics[trim=0.4cm 0.18cm 0.1cm 0.41cm, clip=true, width=4.1cm]{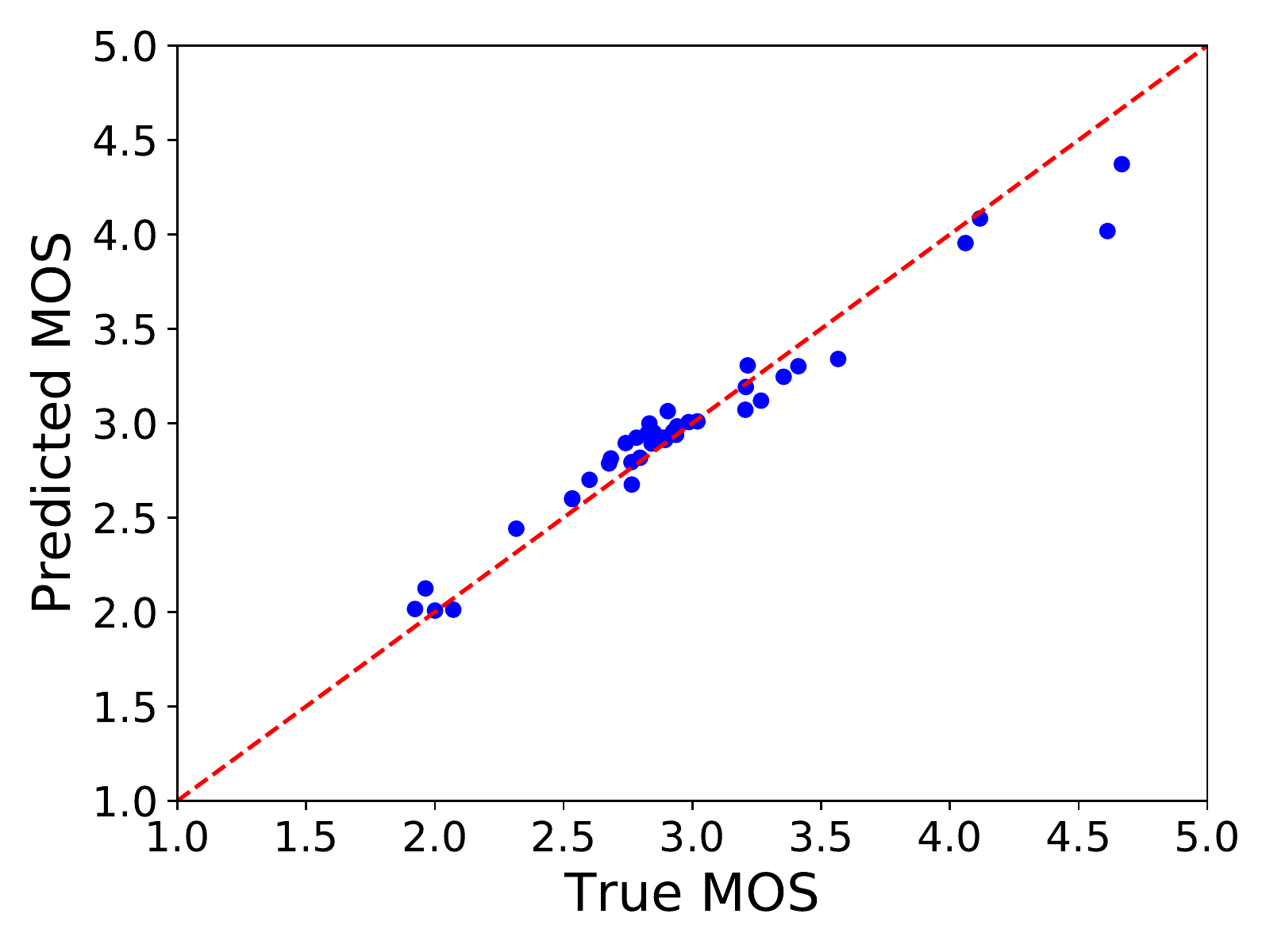} &
            \includegraphics[trim=0.2cm 0.18cm 0.3cm 0.41cm, clip=true, width=4.1cm]{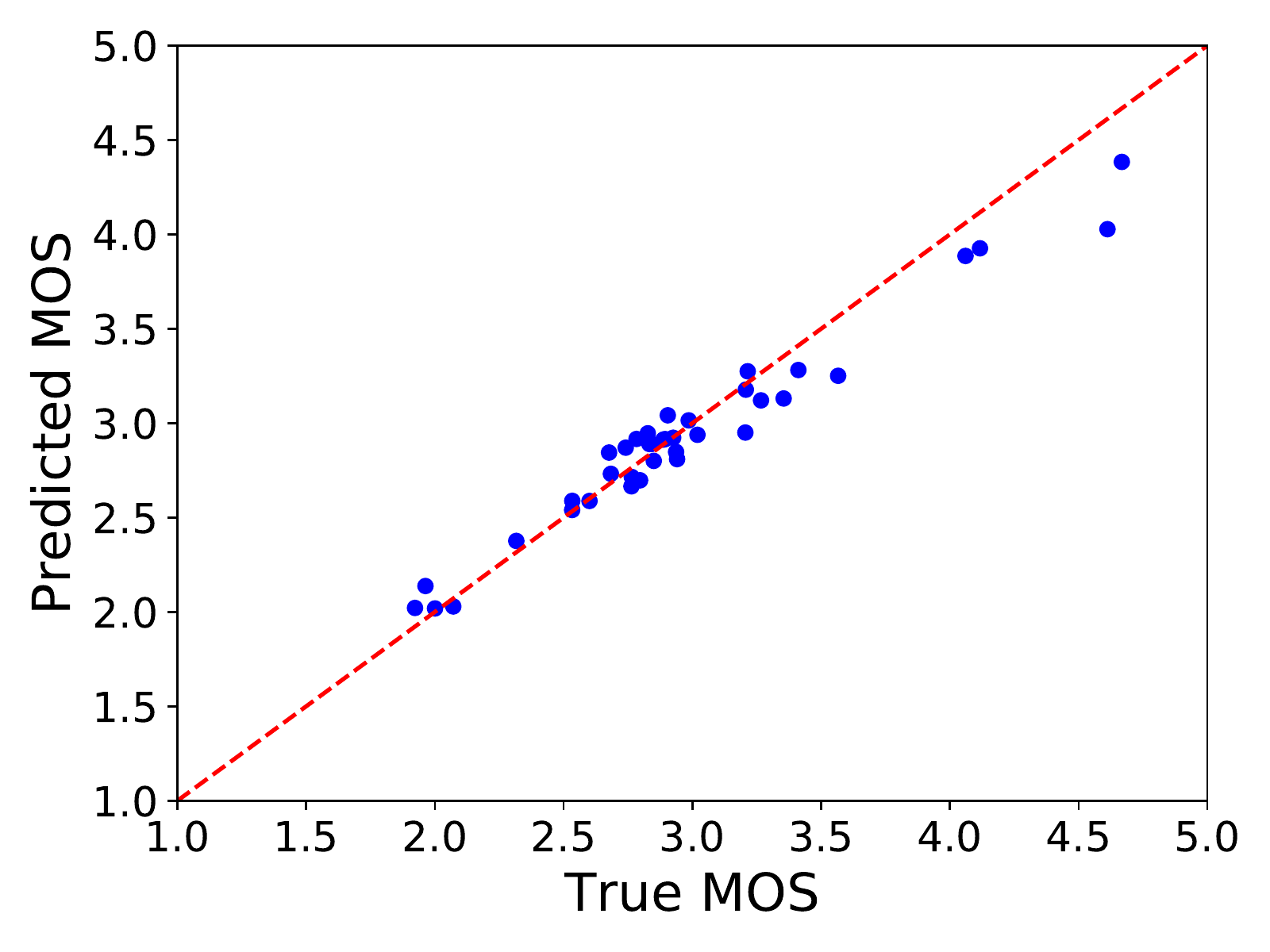}  \\
            (c) +STC & (d) +STC +SD
        \end{tabular}
    \end{small}
%\vspace{-0.25cm}
\caption{
Scatter plots of system-level MOSs for (a) Baseline, (b) MTL model with SD (+SD), (c) MTL model with STC (+STC), and (d) MTL model with both tasks (+STC +SD).
}
%\vspace{0.5cm}
\label{fig:scatter_plot}
\end{figure}

Fig. \ref{fig:scatter_plot} shows the scatter plots of system-level MOSs for the four models. Each dot corresponds to an individual system. 
Comparing (a) and (b), we can see that the systems with high MOSs are better aligned using MTL with SD. % Comparing (a) and (b), the systems with high MOSs are better aligned in (b) (i.e., +SD). 
From this result, it appears that using SD is useful for MOS prediction for high MOSs, as discussed in Section \ref{sec:auxiliary}.
%This indicates that using SD is useful for MOS prediction in high MOS, as discussed in Section \ref{sec:auxiliary}.
Comparing (a) and (c), we can see that the dots more tend to be aligned along the $y = x$ line (red dashed line) when using MTL with STC, which means that the predicted MOSs are more close to ground-truth MOSs. 
%Comparing (a) and (c), we can see that most of the dots get closer to the $y = x$ line (red dashed line) when using MTL with STC, which means that the predicted MOSs are more close to ground-truth MOSs. 
The scatter plot of system-level MOSs for +STC +SD, shown in (d), looks like the combination of (b) and (c).

\subsection{Generalization to unseen spoofing types and raters}
\label{sec:robustness}

As mentioned in the introduction, inter-rater variability exists in MOSs.
Different raters evaluate speech quality differently based on their subjective criteria without objective criteria, leading to high variance within the evaluated scores of the same speech. %Different raters evaluate speech quality based on their own subjectivity without objective criteria, leading to high variance within the evaluated scores of the same speech \cite{MOSNet}. 
This variability caused by the raters fundamentally limits the generalization ability of a MOS prediction model.
%There are two similar spoofing systems between those of VCC'16 and VCC'18. However, they are different.
%Furthermore, the source and target speakers do not overlap between the VCC'16 and VCC'18 data.
%As the number of similar spoofing systems is negligibly small compared to the total number of spoofing types (i.e. spoofing systems and human speakers), we can see the generalization ability of the models to unseen spoofing types and raters from the test results on VCC'16.
%Therefore, we can say that the number of similar spoofing systems is negligibly small compared to the total number of spoofing types (i.e. spoofing systems and human speakers).
As spoofing types and raters do not overlap between the VCC'16 and VCC'18 data, 
we can see the generalization ability of the models to unseen spoofing types and raters from the test results on VCC'16.
The results are shown in Table \ref{tab:mos_results}.

+SD shows worse performance than that of the baseline on the VCC'16 data. 
We conjecture that it is because the average system-level MOS of VCC'16 (i.e., 2.73) is lower than that of VCC'18 (i.e., 2.97) and thus VCC'16 is more vulnerable to the \emph{ignoring effect}.
%+SD shows worse performance than that of the baseline on the VCC'16 data, and we conjecture that it is due to the deficiency of low-level information caused by the \emph{ignoring effect}.
%Furthermore, the average system-level MOS is much lower in VCC'16 than in VCC'18, which leads to be more vulnerable by larger \emph{ignoring effect}.
Using STC together achieves an average relative improvement of 18.4\% over +SD, in terms of three metrics on VCC'16.
Meanwhile, +STC by itself shows better generalization ability than the baseline. 
%Meanwhile, we can see that +STC shows better generalization ability than the baseline on VCC'16. 
It implies that, in the process of distinguishing various systems, the shared layers learn useful low-level features that can be generalized across unseen spoofing types and raters.
%We argue that it is because, in the process of distinguishing various systems, the shared layers learn useful low-level features that can be generalized to unseen spoofing types and raters.
+STC +SD outperforms +STC on the VCC'16 data.
%Moreover, +STC +SD achieves a relative improvement of 10\% and 55\% over +STC and +SD, respectively, on the VCC'16 data.
Based on this result, we infer that the \emph{ignoring effect} of SD can act as a regularizer.
It regularizes +STC by diminishing the \emph{distinguishing effect} of STC, thus encouraging the generalization of the model to unseen spoofing types and raters.

\begin{figure*}[t]
    \centering
    \includegraphics[trim=0.15cm 10.7cm 0.1cm 0.3cm, clip=true, width=17.6cm]{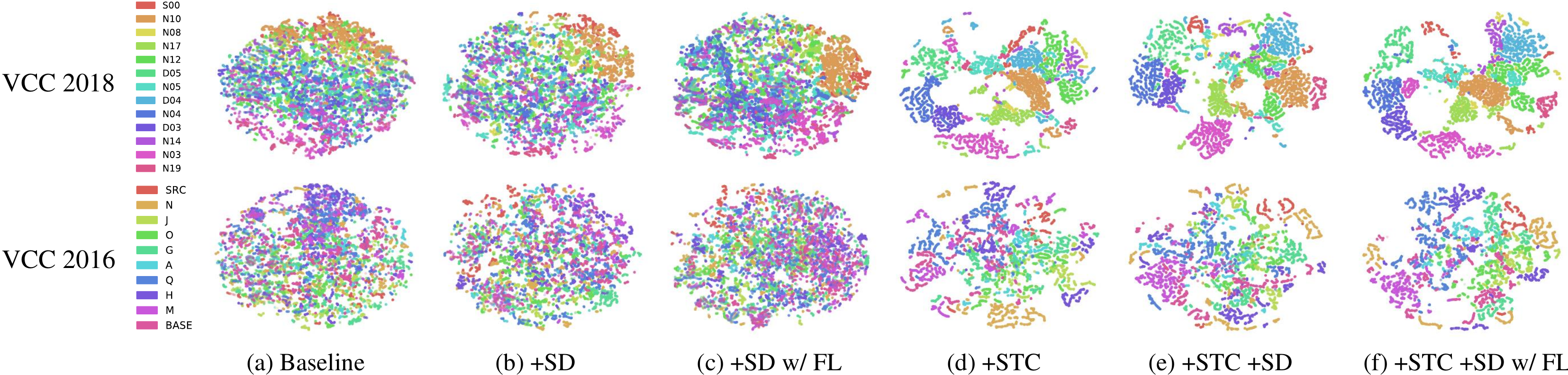}
%\vspace{-0.2cm}
\caption{t-SNE plots of shared features of the models. The legends are sorted in descending order based on system-level MOSs.}
\label{fig:tsne}
\end{figure*}

\subsection{Effect of the focal loss for spoofing detection}
\label{sec:result_FL_SD}
In this section, we discuss the effect of the focal loss (FL) based on the comparison between the second and third rows, and between the fifth and sixth rows of Table \ref{tab:mos_results}.
As discussed in Section \ref{sec:FL_SD}, if we use the FL for SD in training, a model tends to keep learning from the utterances with high SD losses, but to stop learning from those that have relatively low SD losses. 
That is, the FL keeps the \emph{concentrating effect} and decreases the \emph{ignoring effect} of SD. 

For VCC'18, +SD w/ FL plays a similar role with +SD in that it helps MOS prediction by the \emph{concentrating effect}.
As discussed in Section \ref{sec:auxiliary_tasks}, the \emph{ignoring effect} does not harm the performance of +SD, and thus the FL is not useful for +SD.
For VCC'16, however, the \emph{ignoring effect} can be harmful because the model is trained on the VCC'18 training set, which has different spoofing types and human raters from the VCC'16 data. 
Thus, the FL improves the generalization ability of +SD by relieving the \emph{ignoring effect}.
%Note that there is no conflict between the \emph{ignoring effect} and \emph{distinguishing effect} since we do not use the STC task in this case.
%Note that the conflict between the \emph{ignoring effect} and \emph{distinguishing effect} is beside the discussion since we do not use STC in this case.
In this case, we do not need to consider the conflict between the \emph{ignoring effect} and \emph{distinguishing effect} since we do not use STC.

When we use STC together with SD, applying the FL to SD improves the performance for VCC'18 by relieving the \emph{ignoring effect}, which has a conflict with the \emph{distinguishing effect} and limits the performance of +STC +SD.
%Applying the FL to +STC +SD improves the performance for VCC'18 by relieving the \emph{ignoring effect}, which has a conflict with the \emph{distinguishing effect} and limits the performance of +STC +SD.
For VCC'16, +STC +SD w/ FL achieves little better than +STC +SD.
We also compare the proposed model (+STC +SD w/ FL) with a state-of-the-art model, MOSNet+EL \cite{Choi}, which is shown in the last row of Table \ref{tab:mos_results}. MOSNet+EL is an improved version of MOSNet that uses the Encoding Layer \cite{DeepTEN} for the aggregation of frame-level scores.
Here, we train MOSNet+EL with the same experimental setup as for the other models.
The results show that our proposed MTL model outperforms the state-of-the-art model and achieves the best relative improvement of 11.6\% compared to MOSNet.
%In Table \ref{tab:SOTA_results}, we compare the proposed model (+STC +SD w/ FL) with state-of-the-art models. The first and second rows indicate MOSNet \cite{MOSNet} and its advanced version, MOSNet+EL (i.e., MOSNet using the Encoding Layer \cite{Choi}), respectively.
%Here, we train MOSNet+EL with the same experimental setup as other models.
%The results show that our proposed MTL model outperforms the state-of-the-art models.
%Specifically, +STC +SD w/ FL achieves the best relative improvement of 11.6\% compared to MOSNet.

\subsection{Visualization of shared features}
\label{sec:visulization}

To further support our analysis using three types of effects, we visualized the shared features using t-distributed stochastic neighbor embedding (t-SNE) in Fig. \ref{fig:tsne}. 
A single dot represents a frame-level shared feature of an utterance and is colored according to the corresponding system.
%We displayed a frame-level shared feature of an utterance as a single dot and colorized it according to the corresponding system.% to which the utterance belongs.

During visualization for VCC'18, we first considered two systems made by one team as one system since they usually had similar acoustic-prosodic characteristics and MOSs. %, resulting in 26 systems.
For each of VCC'18 and VCC'16,
%Then for both test data,
we sorted the systems in descending order according to the system-level MOS and evenly selected half of them.
The red indicates the source speaker (`S00' or `SRC'), and the orange indicates the spoofing system with the highest MOS (`N10' or `N').
We randomly selected 390 and 300 utterances among the VCC'18 and VCC'16 data, respectively, so that each system has an average of 15 utterances.

Comparing (b) with (a), we can observe the \emph{concentrating effect} and \emph{ignoring effect} of SD. In (b), human speech frames (i.e., red dots) are more exposed, and high-quality speech frames (i.e., red and orange dots) are farther from the others. 
This implies that +SD not only succeeds in classifying human and synthesized speech but also succeeds in predicting the MOS of high-quality synthesized speech. Therefore, we can confirm the \emph{concentrating effect}.
The low-quality synthesized speech frames (i.e., blue, purple, and pink dots) tend to be more mingled with other synthesized speech frames.
This indicates the \emph{ignoring effect}, which is relieved in (c) by the FL. 
Finally, (d), (e), and (f) clearly shows the \emph{distinguishing effect} of STC.
Compared with (a), (b), and (c), the dots are more clustered according to their systems (i.e., colors).

\section{Conclusion}
\label{sec:conclusion}
We proposed the MTL method to improve neural MOS prediction with two auxiliary tasks: SD and STC.
Besides, we adopt the FL to maximize the synergy between the two tasks.
We also presented a detailed analysis of our approach by introducing three types of effects of the auxiliary tasks. 
With experimental results on the VCC'18 and VCC'16 data, we showed that both the SD and STC tasks improve MOS prediction.
Using the proposed MTL model, we can automatically evaluate and compare multiple speech generation systems.
%Using the proposed MTL model, we can make the evaluation and comparison process of multiple speech generation systems fully automatic.
%The proposed MTL model can automatically evaluate multiple speech generation systems, and thus we can compare them.
%The proposed MTL model can automatically evaluate and compare multiple speech generation systems.
%The proposed MTL model can be used to automatically evaluate and compare multiple speech generation systems.
For future work, we will consider the correlation coefficients as well as the MSE in the loss function for MOS prediction. %extend our approach to increase the SRCC, which is always lower than the LCC. 
%For future work, we will consider the ranks of the utterance-level MOSs to increase the SRCC. 
Moreover, using the MOS predicted by our proposed model, we will directly guide a speech generation model to synthesize speech with high MOS as well as low MSE.

\bibliographystyle{IEEEbib}
\bibliography{template}

\begin{thebibliography}{10}

\bibitem{WaveNet}
A.~Oord, S.~Dieleman, H.~Zen, K.~Simonyan, O.~Vinyals, A.~Graves,
  N.~Kalchbrenner, A.~Senior, and K.~Kavukcuoglu,
\newblock ``Wave{N}et: A generative model for raw audio,''
\newblock {\em arXiv preprint arXiv:1609.03499}, 2016.

\bibitem{Tacotron2}
J.~Shen, R.~Pang, R.~J. Weiss, M.~Schuster, N.~Jaitly, Z.~Yang, Z.~Chen,
  Y.~Zhang, Y.~Wang, R.~Skerry-Ryan, R.~A. Saurous, Y.~Agiomyrgiannakis, and
  Y.~Wu,
\newblock ``Natural {TTS} synthesis by conditioning {WaveNet} on {M}el
  spectrogram predictions,''
\newblock in {\em Proc. of the IEEE International Conference on Acoustics,
  Speech and Signal Processing (ICASSP)}, 2018, pp. 4779--4783.

\bibitem{DeepVoice3}
W.~Ping, K.~Peng, A.~Gibiansky, S.~O. Arik, A.~Kannan, S.~Narang, J.~Raiman,
  and J.~Miller,
\newblock ``Deep {V}oice 3: Scaling text-to-speech with convolutional sequence
  learning,''
\newblock in {\em Proc. of the International Conference on Learning
  Representations (ICLR)}, 2018.

\bibitem{TransformerTTS}
N.~Li, S.~Liu, Y.~Liu, S.~Zhao, and M.~Liu,
\newblock ``Neural speech synthesis with {T}ransformer network,''
\newblock in {\em Proc. of the AAAI Conference on Artificial Intelligence},
  2019, pp. 6706--6713.

\bibitem{CycleGAN}
T.~Kaneko and H.~Kameoka,
\newblock ``Cycle{GAN-VC}: Non-parallel voice conversion using cycle-consistent
  adversarial networks,''
\newblock in {\em Proc. of 2018 26th European Signal Processing Conference
  (EUSIPCO)}, 2018, pp. 2100--2104.

\bibitem{NeuralTTS_VC}
Z.~Kons, S.~Shechtman, A.~Sorin, R.~Hoory, C.~Rabinovitz, and E.~D.~S. Morais,
\newblock ``Neural {TTS} voice conversion,''
\newblock in {\em Proc. of Spoken Language Technology Workshop (SLT)}, 2018,
  pp. 290--296.

\bibitem{LimitOfMOS}
M.~Chu and H.~Peng,
\newblock ``An objective measure for estimating {MOS} of synthesized speech,''
\newblock in {\em Proc. of Eurospeech}, 2001.

\bibitem{Eval}
P.~Wagner, J.~Beskow, S.~Betz, J.~Edlund, J.~Gustafson, G.~E. Henter, S.~L.
  Maguer, Z.~Malisz, E.~Sz{\'{e}}kely, C.~T{\aa}nnander, and J.~Vo{\ss}e,
\newblock ``Speech synthesis evaluation — state-of-the-art assessment and
  suggestion for a novel research program,''
\newblock in {\em Proc. 10th Speech Synthesis Workshop (SSW10)}, 2019.

\bibitem{MCD}
R.~Kubichek,
\newblock ``Mel-cepstral distance measure for objective speech quality
  assessment,''
\newblock in {\em Proc. IEEE Pacific Rim Conference on Communications Computers
  and Signal Processing}, 1993, pp. 125--128.

\bibitem{PESQ}
A.~W. Rix, J.~G. Beerends, M.~P. Hollier, and A.~P. Hekstra,
\newblock ``Perceptual evaluation of speech quality ({PESQ})-a new method for
  speech quality assessment of telephone networks and codecs,''
\newblock in {\em Proc. of the IEEE International Conference on Acoustics,
  Speech and Signal Processing (ICASSP)}, 2001, pp. 749--752.

\bibitem{ANIQUE}
D.-S. Kim,
\newblock ``{ANIQUE}: an auditory model for single-ended speech quality
  estimation,''
\newblock {\em IEEE Transactions on Speech and Audio Processing}, vol. 13, no.
  5, pp. 821--831, 2005.

\bibitem{P563}
L.~Malfait, J.~Berger, and M.~Kastner,
\newblock ``P. 563: The {ITU-T} standard for single-ended speech quality
  assessment,''
\newblock {\em IEEE Transactions on Audio, Speech, and Language Processing},
  vol. 14, no. 6, pp. 1924--1934, 2006.

\bibitem{Choi}
Y.~Choi, Y.~Jung, and H.~Kim,
\newblock ``Deep {MOS} predictor for synthetic speech using cluster-based
  modeling,''
\newblock in {\em Proc. of Interspeech}, 2020, pp. 1743--1747.

\bibitem{HierarchicalAssessment}
T.~Yoshimura, G.~E. Henter, O.~Watts, M.~Wester, J.~Yamagishi, and K.~Tokuda,
\newblock ``A hierarchical predictor of synthetic speech naturalness using
  neural networks,''
\newblock in {\em Proc. of Interspeech}, 2016, pp. 342--346.

\bibitem{AutoMOS}
B.~Patton, Y.~Agiomyrgiannakis, M.~Terry, K.~W. Wilson, R.~A. Saurous, and
  D.~Sculley,
\newblock ``Auto{MOS}: Learning a non-intrusive assessor of
  naturalness-of-speech,''
\newblock in {\em Proc. of NIPS End-to-end Learning for Speech and Audio
  Processing Workshop}, 2016.

\bibitem{MOSNet}
C.~Lo, S.~Fu, W.~Huang, X.~Wang, J.~Yamagishi, Y.~Tsao, and H.~Wang,
\newblock ``{MOSNet}: Deep learning based objective assessment for voice
  conversion,''
\newblock in {\em Proc. of Interspeech}, 2019, pp. 1541--1545.

\bibitem{MTL_1993}
R.~A. Caruana,
\newblock ``Multitask learning: A knowledge-based source of inductive bias,''
\newblock in {\em Proc. of the international conference on machine learning
  (ICML)}, 1993.

\bibitem{MTL_overview}
S.~Ruder,
\newblock ``An overview of multi-task learning in deep neural networks,''
\newblock {\em arXiv preprint arXiv:1706.05098}, 2017.

\bibitem{MTL_NLP}
S.~Subramanian, A.~Trischler, Y.~Bengio, and C.~J. Pal,
\newblock ``Learning general purpose distributed sentence representations via
  large scale multi-task learning,''
\newblock in {\em Proc. of the International Conference on Learning
  Representations (ICLR)}, 2018.

\bibitem{MTL_SR}
S.~Toshniwal, H.~Tang, L.~Lu, and K.~Livescu,
\newblock ``Multitask learning with low-level auxiliary tasks for
  encoder-decoder based speech recognition,''
\newblock in {\em Proc. of Interspeech}, 2017.

\bibitem{MTL_KD}
T.~Maekaku, Y.~Kida, and A.~Sugiyama,
\newblock ``Simultaneous detection and localization of a wake-up word using
  multi-task learning of the duration and endpoint,''
\newblock in {\em Proc. of Interspeech}, 2019, pp. 4240--4244.

\bibitem{MTL_SV_high_order}
L.~You, W.~Guo, L.~Dai, and J.~Du,
\newblock ``Multi-task learning with high-order statistics for x-vector based
  text-independent speaker verification,''
\newblock in {\em Proc. of Interspeech}, 2019, pp. 1158--1162.

\bibitem{MTL_SV_noisy}
A.~Jati, R.~Peri, M.~Pal, T.~J. Park, N.~Kumar, R.~Travadi, P.~Georgiou, and
  S.~Narayanan,
\newblock ``Multi-task training of hybrid {DNN-TVM} model for speaker
  verification with noisy and far-field speech,''
\newblock in {\em Proc. of Interspeech}, 2019, pp. 2463--2467.

\bibitem{BeneficialTask}
J.~Bingel and A.~S{\o}ggard,
\newblock ``Identifying beneficial task relations for multi-task learning in
  deep neural networks,''
\newblock {\em arXiv preprint arXiv:1702.08303}, 2017.

\bibitem{NER_aux}
S.~Louvan and B.~Magnini,
\newblock ``Exploring named entity recognition as an auxiliary task for slot
  filling in conversational language understanding,''
\newblock in {\em Proc. of the 2018 EMNLP Workshop on Search-Oriented
  Conversational AI (SCAI)}, 2018, pp. 74--80.

\bibitem{FocalLoss}
T.~Y. Lin, P.~Goyal, R.~Girshick, K.~He, and P.~Dollár,
\newblock ``Focal loss for dense object detection,''
\newblock in {\em Proc. of the IEEE international conference on computer vision
  (ICCV)}, 2017, pp. 2980--2988.

\bibitem{VCC2018}
J.~Lorenzo-Trueba, J.~Yamagishi, T.~Toda, D.~Saito, F.~Villavicencio,
  T.~Kinnunen, and Z.~Ling,
\newblock ``The voice conversion challenge 2018: Promoting development of
  parallel and nonparallel methods,''
\newblock in {\em Proc. of Odyssey The Speaker and Language Recognition
  Workshop}, 2018, pp. 195--202.

\bibitem{SSD}
P.~L. De~Leon, I.~Hernaez, I.~Saratxaga, M.~Pucher, and J.~Yamagishi,
\newblock ``Detection of synthetic speech for the problem of imposture,''
\newblock in {\em Proc. of the IEEE International Conference on Acoustics,
  Speech and Signal Processing (ICASSP)}, 2011, pp. 4844--4847.

\bibitem{bootstrap}
B.~Efron and R.~J. Tibshirani,
\newblock {\em An Introduction to the Bootstrap},
\newblock Chapman \& Hall/CRC, 1994.

\bibitem{Pearson}
K.~Pearson,
\newblock ``Notes on the history of correlation,''
\newblock {\em Biometrika}, vol. 13, no. 1, pp. 25--45, 1920.

\bibitem{Spearman}
C.~Spearman,
\newblock ``The proof and measurement of association between two things,''
\newblock {\em The American Journal of Psychology}, vol. 15, no. 1, pp.
  72--101, 1904.

\bibitem{VCC2016}
T.~Toda, L.~Chen, D.~Saito, F.~Villavicencio, M.~Wester, Z.~Wu, and
  J.~Yamagishi,
\newblock ``The {V}oice {C}onversion {C}hallenge 2016,''
\newblock in {\em Proc. of Interspeech}, 2016, pp. 1632--1636.

\bibitem{DeepTEN}
H.~Zhang, J.~Xue, and K.~Dana,
\newblock ``Deep {TEN}: Texture encoding network,''
\newblock in {\em Proc. of Computer Vision and Pattern Recognition (CVPR)},
  2017, pp. 708--717.

\end{thebibliography}
\balance

\end{document}